\documentstyle[prl,aps,preprint,floats]{revtex}

\begin{document}
\preprint{ESI-1413}

\title{Two-dimensional metric-affine gravity}

\author{Yuri N.~Obukhov\footnote{On leave from: Dept. of Theoret. 
Physics, Moscow State University, 117234 Moscow, Russia}}
\address{Institute for Theoretical Physics, University of Cologne,
50923 K\"oln, Germany}

\maketitle

\begin{abstract}
There is a number of completely integrable gravity theories in two dimensions.
We study the metric-affine approach on a 2-dimensional spacetime and display a
new integrable model. Its properties are described and compared with the known
results of Poincar\'e gauge gravity. 
\end{abstract}
\bigskip


\section{Introduction}

Although it is well known that the Hilbert-Einstein Lagrangian yields the 
trivial dynamics in 2 dimensions, 2-dimensional gravitational theory
has attracted considerable attention recently. In particular, the extension
of the spacetime geometry by including nontrivial torsion has encompassed 
a class of interesting models with rich mathematical and physical contents
\cite{kat,kum,str,sol,col}. For an overview of the relevant results and a
more exhaustive list of publications, we refer to \cite{rev1,rev2}. These
models are remarkable for at least two reasons. Namely, they provide a 
convenient tool for the study of sting-motivated dilaton gravity which
can be formulated as an effective Poincar\'e gauge model. And moreover, 
they represent several examples of physically meaningful completely 
integrable models. 

The purpose of this paper is to generalize the 2-dimensional gravity models by 
taking into account all possible post-Riemannian geometrical structures. In 
practical terms, this means to take into consideration both the torsion and
the nonmetricity, thus extending the earlier results \cite{kat,kum,str,sol,col}
which were confined only to metric-compatible connections with torsion.
The general framework for metric-affine gravity (MAG) is firmly established
on the basis of the gauge approach for the general affine group in a spacetime
of any dimension \cite{PR}. The specific applications to string and dilaton
gravity were discussed in \cite{dil}, for example. The first (to our knowledge)
2-dimensional gravity model with nonmetricity \cite{tekin}, however, was 
constructed in terms of a nonmetric geometry with vanishing torsion. It is thus
of interest to study the case of the most general metric-affine 2-dimensional 
spacetime with both torsion and nonmetricity being nontrivial. Technically,
it seems natural to start with the quadratic Poincar\'e gauge model and to 
extend it by including the new quadratic nonmetricity term in the Lagrangian. 
We then will benefit from the techniques developed previously for the 
Poincar\'e gauge case \cite{col,rev1}. Our main result is the demonstration 
of the complete integrability of the obtained model in vacuum. 

The structure of the paper is as follows. In Sec.~\ref{magst} we briefly
summarize the basics of the MAG approach, with special attention to the
irreducible decomposition of the geometric objects in 2 dimensions. 
Sec.~\ref{TTds} describes how the spacetime interval can be constructed
with the help of the torsion, whereas Sec.~\ref{eqs} is devoted to the
formulation of the dynamical scheme of the model. The complete integrability
in vacuum is demonstrated in Sec.~\ref{sol}. Finally, in Sec~\ref{disc} we
discuss the results obtained and outline the open problems. 

Our basic notations and conventions are those of \cite{PR}. In particular,
the signature of the 2-dimensional metric is assumed to be $(-,+)$. Spacetime 
coordinates are labeled by Latin indices, $i,j,\dots = 0,1$ (for example, 
$dx^i$), whereas Greek indices, $\alpha,\beta,\dots = 0,1$, label the local 
frame components (for example, the coframe 1-form $\vartheta^\alpha$). Along 
with the coframe 1-forms $\vartheta^{\alpha}$, we will use the so called 
$\eta$-basis of the dual coframes. Namely, we define the Hodge dual such that 
$\eta:={}^*1$ is the volume 2-form. Furthermore, denoting by $e_{\alpha}$ the 
vector frame, we have the 1-form $\eta_\alpha := e_\alpha\rfloor\eta = 
{}^*\vartheta_\alpha$, and the 0-form $\eta_{\alpha\beta}:= e_\beta\rfloor
\eta_\alpha ={}^*(\vartheta_\alpha\wedge\vartheta_\beta)$. The last expression
represents the 2-dimensional totally antisymmetric Levi-Civita tensor.

\section{Metric-affine approach}\label{magst}

In metric-affine gravity (MAG), the gravitational field is described
by the coframe one-form $\vartheta^\alpha$, the linear connection one-form
$\Gamma_\alpha{}^\beta$ and the metric $g_{\alpha\beta}$. The first two 
variables are considered to be the gauge potentials of the gravitational 
field corresponding, respectively, to the translation group and and the 
general linear group acting in the tangent space at each point of spacetime. 
The gravitational field strengths are given by the torsion two-form $T^\alpha 
:= D\vartheta^\alpha$, the curvature two-form $R_\alpha{}^\beta:= 
d\Gamma_\alpha{}^\beta -\Gamma_\alpha{}^\gamma\wedge\Gamma_\gamma{}^\beta$,
and the nonmetricity 1-form $Q_{\alpha\beta} := - Dg_{\alpha\beta}$. The 
frame $e_{\alpha} = e^{i}{}_{\alpha}\,\partial_{i}$ is dual to the coframe 
$\vartheta^{\beta} = e_{j}{}^{\beta}\,dx^{j}$, i.e., $e_{\alpha}\rfloor 
\vartheta^\beta = e^i{}_{\alpha}\,e_i{}^\beta = \delta_\alpha^\beta$. The 
spacetime manifold $M$ is equipped with a line element
\begin{equation}
ds^2 = g_{ij}\,dx^{i}\otimes dx^{j} = g_{\alpha\beta}\,\vartheta^\alpha
\otimes\vartheta^\beta.\label{interval}
\end{equation}
More details about the MAG approach in any dimension can be found in \cite{PR}.


Before we proceed with the analysis of the metric-affine gravity model in 
two dimensions, it is instructive to understand the structure of the 
basic geometric objects. Here we describe their independent components
and irreducible parts.

In two dimensions, the {\it torsion} has 2 components. It reduces to its vector
trace 1-form (second irreducible piece \cite{PR})
\begin{equation}
T^\alpha = \vartheta^{\alpha}\wedge T,\qquad T:= e_\alpha\rfloor T^\alpha.
\end{equation} 

The 2-dimensional {\it nonmetricity} has 6 independent components and it
decomposes into 3 irreducible parts:
$Q_{\alpha\beta}={}^{(1)}Q_{\alpha\beta}+
{}^{(3)}Q_{\alpha\beta}+{}^{(4)}Q_{\alpha\beta}$, with 
\begin{eqnarray}
{}^{(3)}Q_{\alpha\beta}&:=& \vartheta_{(\alpha}e_{\beta)}
\rfloor\Lambda - {1\over 2}g_{\alpha\beta}\Lambda,\label{Q3}\\
{}^{(4)}Q_{\alpha\beta}&:=&g_{\alpha\beta}Q,\label{Q4}\\
{}^{(1)}Q_{\alpha\beta}&:=&Q_{\alpha\beta} - {}^{(3)}Q_{\alpha\beta}
- {}^{(4)}Q_{\alpha\beta}.\label{Q1}
\end{eqnarray}
Here the shear covector part of the nonmetricity and the Weyl covector are, 
respectively,
\begin{equation}
\Lambda:=\vartheta^{\alpha}e^{\beta}\rfloor
{\nearrow\!\!\!\!\!\!\!Q}_{\alpha\beta},\quad\quad
Q:={1\over 2}g^{\alpha\beta}Q_{\alpha\beta},
\end{equation}
where ${\nearrow\!\!\!\!\!\!\!Q}_{\alpha\beta}=Q_{\alpha\beta}- Q 
g_{\alpha\beta}$ is the traceless piece of the nonmetricity. For the 
spacetime dimension greater than 2, the nonmetricity has also a second
irreducible piece \cite{PR} which vanishes identically ${}^{(2)}
Q_{\alpha\beta} = 0$ in 2 dimensions. 

The three 1-forms of torsion and nonmetricity $(T,\,Q,\,\Lambda )$
form the so-called {\it triplet} which plays a significant role in the 
MAG theories in 4 dimensions \cite{eff}. As we will see, the triplet of
1-forms is important also for understanding of the 2-dimensional MAG.

The {\it curvature} 2-form has 4 components in 2 dimensions and it 
decomposes into the 3 irreducible pieces:
\begin{equation}
R_{\alpha\beta} = W_{\alpha\beta} + {\nearrow\!\!\!\!\!\!\!Z}_{\alpha\beta} 
+ {\frac 12}\,g_{\alpha\beta}\,Z.
\end{equation}
Here the skew-symmetric part $W_{\alpha\beta} := R_{[\alpha\beta]}$ is the 
direct generalization of the Riemann-Cartan curvature, whereas the symmetric
part $Z_{\alpha\beta} := R_{(\alpha\beta)} = {\frac 12}\,DQ_{\alpha\beta}$ is 
only nontrivial in presence of the nonmetricity. The skew-symmetric part is 
irreducible in 2 dimensions. It has one independent component and it can 
be expressed in terms of the curvature scalar $R:=e_\alpha\rfloor e_\beta
\rfloor R^{\alpha\beta}$:
\begin{equation}
W^{\alpha\beta}= -\,{1\over 2}\,R\;\vartheta^{\alpha}\wedge\vartheta^{\beta}.
\end{equation} 
However the symmetric part is more nontrivial. It has 3 components and it 
can be decomposed into the trace 2-form $Z := g^{\alpha\beta}Z_{\alpha\beta}$ 
(1 component) and the traceless part ${\nearrow\!\!\!\!\!\!\!Z}_{\alpha\beta}
:= Z_{\alpha\beta} - {\frac 12}g_{\alpha\beta}\,Z$ (2 components).

The coframe $\vartheta^\alpha$ is not covariantly constant in a general MAG
spacetime. Similarly, the covariant derivatives of the $\eta$-objects are
also non-vanishing. In particular, in 2 dimensions, we find explicitly:
\begin{eqnarray}
D\eta^\alpha &=& \eta\,e^\alpha\rfloor\Lambda 
+ \eta^{\alpha\beta}\,T_\beta,\label{Da}\\
D\eta^\alpha{}_\beta &=& {}^\ast\left({}^{(1)}Q^\alpha{}_\beta
- {}^{(3)}Q^\alpha{}_\beta\right). \label{Dab}
\end{eqnarray}
We will use these identities in the analysis of the MAG field equations.

\section{Torsion and line element of spacetime}\label{TTds}

In our demonstration of the complete integrability of the 2-dimensional MAG
model, we will use the technical tools developed earlier for the Poincar\'e
gauge theory \cite{col,rev1}. Namely, in 2 dimensions, one can construct 
the spacetime interval from a non-degenerate torsion. Here we briefly 
summarize the corresponding definitions and results. 

In 2 dimensions, the two independent components of the torsion can be 
described in terms of the vector-valued torsion zero-form  $t^\alpha$ defined
via the Hodge dual
\begin{equation}
  t^{\alpha} := {}^\ast T^{\alpha}.
\end{equation}
Then the torsion 2-form is recovered as
\begin{equation}
  T^{\alpha} = - t^{\alpha}\eta.
\end{equation} 
We call the manifold $M$ a {\it non-degenerate metric-affine spacetime}
when the torsion square is not identically zero, i.e. $t^2 := t_{\alpha}
\,t^{\alpha}\neq 0$. Then we can write a coframe as \cite{col,rev1}:
\begin{equation}
\vartheta^{\alpha}= -\,{\frac 1 {t^2}}\left(T\,\eta^{\alpha\beta}\,t_{\beta}
+ {}^\ast T\,t^{\alpha}\right).\label{theta}
\end{equation}
In other words, the torsion 1-form $T$ and its dual ${}^\ast T$ specify a 
coframe with respect to which one can expand all the 2D geometrical objects. 
Such a coframe is non-degenerate when $t^2\neq 0$, hence the terminology. 
The volume 2-form can be calculated as an exterior square of the torsion 
1-form $T$:
\begin{equation}
\eta = {\frac 12}\,\eta_{\alpha\beta}\,\vartheta^{\alpha}\wedge
\vartheta^{\beta}={\frac 1 {t^2}}\,{}^\ast T\wedge T.\label{vol}
\end{equation}
Defining a coframe of a 2-dimensional spacetime in terms of the torsion 1-form 
is an important step in the study of the integrability of the MAG model.

\section{Gravitational field equations}\label{eqs}

We are now ready to formulate the dynamics of the MAG model in two dimensions.
Let us consider the Lagrangian 2-form:
\begin{equation}
V(\vartheta^{\alpha}, T^{\alpha}, R_\alpha{}^\beta, Q_{\alpha\beta}) =
\left({\frac 12}\,R - {\frac b4}\,R^2 - \lambda\right)\eta  
- {\frac {a_1}2}\,T^\alpha\wedge{}^\ast T_\alpha - {\frac {a_2}2}
\,Q^{\alpha\beta}\wedge{}^\ast Q_{\alpha\beta}.\label{V}
\end{equation}
Here $a_1, a_2, b$ and $\lambda$ (cosmological term) are the coupling
constants. Recall that the Lagrangian of the Poincar\'e gauge model in 2 
dimensions contained only the linear and quadratic contractions of 
curvature and torsion -- the first and the second terms on the right-hand
side of (\ref{V}). In order to investigate the influence of the nonmetricity, 
we have only minimally modified the Poincar\'e Lagrangian by adding the last
term quadratic in the nonmetricity. However, it is necessary also to note
that the curvature scalar $R$ is now qualitatively different since it depends
on the general linear connection and not on the Lorentz connection alone.

The gravitational field equations are derived from the total Lagrangian $V + 
L_{\rm mat}$ by independent variations with respect to the metric $g_{\alpha
\beta}$, the 1-form $\vartheta^{\alpha}$ (coframe) and the 1-form $\Gamma_{
\beta}{}^{\alpha}$ (connection). The corresponding so-called {\it zeroth}, 
{\it first} and {\it second} field equations read 
\begin{eqnarray}
DM^{\alpha\beta} - m^{\alpha\beta} &=& \sigma^{\alpha\beta},\label{zeroth}\\
DH_{\alpha}- E_{\alpha}&=&\Sigma_{\alpha}\,,\label{first}\\ 
DH^{\alpha}{}_{\beta}-E^{\alpha}{}_{\beta}&=&\Delta^{\alpha}{}_{\beta}\,.
\label{second}
\end{eqnarray} 
The source terms in the right-hand sides are defined as the derivatives 
of the matter Lagrangian: $\sigma^{\alpha\beta} := 2\delta 
L_{\rm mat}/\delta g_{\alpha\beta}$ (the metrical energy-momentum 2-form), 
$\Sigma_\alpha := \delta L_{\rm mat}/\delta \vartheta^\alpha$ (the canonical 
energy-momentum 1-form), $\Delta^\alpha{}_\beta := \delta L_{\rm mat}/\delta 
\Gamma_\alpha{}^\beta$ (the canonical hypermomentum 1-form). The gauge field 
momenta appearing in the left-hand sides of the field equations are given by
\begin{eqnarray}
M^{\alpha\beta} &:=& -\,2{\partial V \over \partial Q_{\alpha\beta}} =
2a_2\,{}^\ast Q^{\alpha\beta}, \label{Mab}\\
H_{\alpha} &:=& -\,{\partial V \over \partial T^{\alpha}} = 
a_1\,t_\alpha,\label{Ha}\\
H^\alpha{}_\beta &:=& -\,{\partial V \over \partial R_\alpha{}^\beta} =
{\frac 12}\,\eta^\alpha{}_\beta\left(1 - b\,R\right).\label{Hab}
\end{eqnarray}
Finally, the energy-momenta and the hypermomentum of the gravitational field
are described by $m^{\alpha\beta} = 2\partial V/\partial g_{\alpha\beta}$ and
\begin{eqnarray}
E_{\alpha} &=& e_{\alpha}\rfloor V + (e_{\alpha}\rfloor T^{\beta})H_\beta 
+ (e_{\alpha}\rfloor R_\beta{}^\gamma) H^\beta{}_\gamma + {\frac 12}
\,(e_{\alpha}\rfloor Q_{\beta\gamma})\wedge M^{\beta\gamma},\label{Ea}\\
E^{\alpha}{}_{\beta} &=& - \vartheta^{\alpha}\wedge H_{\beta} - 
M^{\alpha}{}_{\beta}.\label{Eab}
\end{eqnarray}

One can show quite generally (in any dimension) \cite{PR} that the zeroth
equation (\ref{zeroth}) is redundant: it is a consequence of (\ref{first}),
(\ref{second}) and of the Noether identities. Accordingly, we will 
consider the system of the first and second field equations (\ref{first})
and (\ref{second}) which determine completely the dynamics of the 
gravitational field. 

\section{General vacuum solution}\label{sol}

We will now specialize to the vacuum case when the matter sources are
absent, $\Sigma_\alpha = 0, \Delta^\alpha{}_\beta = 0$. As a preliminary
remark, we notice that if we put the nonmetricity equal zero at this stage,
$Q_{\alpha\beta} =0$, we will not recover the results of the Poincar\'e gauge
model. Similarly, the limit of $a_2 = 0$ does not yield the old results 
despite the fact that the gravitational Lagrangian then formally coincides 
with that of the 2-dimensional Poincar\'e gauge theory. It is important to 
realize that the MAG dynamics is very different from the Poincar\'e gauge 
case, in particular, the number of the field equations is now greater.

\subsection{Second field equation}

Substituting (\ref{Mab})-(\ref{Hab}) into the second field equation 
(\ref{second}), we find in vacuum:
\begin{equation}
{\frac 12}\,D\left[\eta^\alpha{}_\beta\,(1 - b\,R)\right] + a_1
\,\vartheta^\alpha t_\beta + 2a_2\,{}^\ast Q^\alpha{}_\beta = 0.\label{sec}
\end{equation}
Taking the trace, we obtain the relation between the torsion and the
Weyl 1-forms:
\begin{equation}
Q = {\frac {a_1}{4a_2}}\,T.\label{QT}
\end{equation}
Now, using the identity (\ref{Dab}), we decompose (\ref{sec}) into the
symmetric and antisymmetric parts:
\begin{eqnarray}
{\frac 12}\,(1 - b\,R)\,{}^\ast\left({}^{(1)}Q_{\alpha\beta}
- {}^{(3)}Q_{\alpha\beta}\right) + a_1\,{}^\ast\!\left(\vartheta_{(\alpha}
e_{\beta)}\rfloor T - g_{\alpha\beta}\,T \right)
+ 2a_2\,{}^\ast Q_{\alpha\beta} &=& 0,\label{sym}\\
-\,{\frac b2}\,\eta_{\alpha\beta}\,dR + a_1\,\vartheta_{[\alpha}t_{\beta]}
&=& 0.\label{ant}
\end{eqnarray}
The symmetric part (\ref{sym}) yields 
\begin{equation}\label{LT}
{}^{(1)}Q_{\alpha\beta} = 0,\qquad \Lambda = {\frac {2a_1}{1 - bR - 4a_2}}\,T.
\end{equation}
As a result, we discover at the end the {\it triplet} structure of the 
torsion-nonmetricity sector when the only nontrivial pieces of the torsion
and the nonmetricity remain the three 1-forms which are proportional to each
other: $\Lambda\sim Q \sim T$. 

Finally, the antisymmetric equation (\ref{ant}) yields 
\begin{equation}
T = -\,{\frac {b}{a_1}}\,dR.\label{TdR}
\end{equation}
This is completely analogous to the corresponding result of the Poincar\'e
gauge model \cite{col,rev1}.

\subsection{First field equation}

We begin the analysis of the first MAG field equation (\ref{first}) by
noticing that substitution of (\ref{V}), (\ref{Ha}), (\ref{Hab}) and 
(\ref{Mab}) in (\ref{Ea}) yields
\begin{equation}
E_\alpha = -\,\widetilde{\cal V}\,\eta_\alpha + {\frac {a_2}2}\left(
{}^\ast Q^{\beta\gamma}\,e_\alpha\rfloor Q_{\beta\gamma} + Q_{\beta\gamma}
\,e_\alpha\rfloor{}^\ast Q^{\beta\gamma}\right).\label{Ea1}
\end{equation}
Here $\widetilde{\cal V} = a_1\,t^2/2 - bR^2/4 + \lambda$. Obviously, the
above 1-form has the properties: 
\begin{equation}
\eta^\alpha\wedge E_\alpha =0,\qquad \vartheta^\alpha\wedge E_\alpha = 
-\,\widetilde{\cal V}\,\eta.\label{vtaE}
\end{equation}
Using the results of the previous subsection, we then find the first MAG
equation explicitly:
\begin{equation}
a_1Dt_\alpha = -\,\widetilde{\cal V}\,\eta_\alpha + a_2\left({}^\ast Q
\,e_\alpha\rfloor Q - Q\,\eta_{\alpha\beta}e^\beta\rfloor Q\right).\label{dt}
\end{equation}
This equation determines the components of the torsion vector. Recall, 
however, that ultimately we are interested in the spacetime interval
(\ref{interval}) which is expressed in terms of the coframe (\ref{theta}).
Thus, we need to find the 1-form $T$ and its dual $^\ast T$ together with
the quadratic invariant of the torsion $t^2 = t_{\alpha}\,t^{\alpha}$.
This is achieved by contracting (\ref{dt}) with $\eta^\alpha$, 
$\vartheta^\alpha$, and $t^\alpha$, respectively. 

Contracting (\ref{dt}) with $\eta^\alpha$ and taking into account (\ref{vtaE}),
we find $d\,T = 0$. Note that we need the identity (\ref{Da}) and the relation
(\ref{LT}) for this. The result is consistent with the earlier 
explicit formula (\ref{TdR}). 

Contracting (\ref{dt}) with $\vartheta^\alpha$, we obtain 
\begin{equation}
a_1\,d\,{}^\ast T = \left(a_1\,t^2 - 2\widetilde{\cal V}\right)\eta.\label{dhT}
\end{equation}
Finally, contracting (\ref{dt}) with $t^\alpha$, we get the differential
equation for the function $t^2$:
\begin{equation}
{\frac {a_1}2}\,dt^2 = \widetilde{\cal V}\,T + {\frac {a_1}4}
\,t^2\,(Q + \Lambda).\label{dt2}
\end{equation}
Substituting (\ref{QT}), (\ref{LT}) and (\ref{TdR}) into (\ref{dt2}), we 
obtain the first-order ordinary differential equation, the integration of 
which yields $t^2$ as a function of the curvature scalar:
\begin{equation}
-\,t^2 = {\frac {a^2\rho}{2a_1^2b}}\Bigg\{{\frac {[4b\lambda - (1 - 4a_2)2]}
{a}}\,e^{-\rho}\,{\rm Ei}(\rho) 
 -\,a\rho + a - 2(1 - 4a_2) + c_0\,e^{-\rho}\Bigg\}.\label{t2}
\end{equation}
Here ${\frac 1 a} := {\frac {1}{a_1}} + {\frac {1}{8a_2}}$, and ${\rm Ei}
(\rho)$ is the exponential integral function of the variable
\begin{equation}
\rho = {\frac {bR + 4a_2 - 1}{a}}.\label{rho}
\end{equation}
The meaning of the integration constant $c_0$ will be clarified in the
next subsection.

\subsection{Spacetime geometry}

In order to find the spacetime interval, we will proceed along the same
line as the in \cite{col,rev1}. Namely, as we see from (\ref{TdR}), the 
torsion 1-form $T$ plays the role of the first leg of a zweibein and we
can interpret $R$ as one of the local spacetime coordinates. The form of
the solution (\ref{t2}) suggests, furthermore, that it will be more convenient 
to replace $R$ with $\rho$ using the linear transformation (\ref{rho}). 

The second leg of a zweibein will be then described by the dual torsion 
1-form $^\ast T$. Then following \cite{col,rev1} we introduce the second
local coordinate $\zeta$ and find in this way the general ansatz
\begin{equation}
{}^\ast T = B(\zeta,\rho)\,d\zeta.\label{hT}
\end{equation}
The unknown function $B(\zeta,\rho)$ is determined as follows. Substituting 
(\ref{hT}) into (\ref{dhT}) and using (\ref{vol}), we find the differential 
equation:
\begin{equation}
{\frac {a_1}a}\partial_\rho B = \left(a_1\,t^2 - 2\widetilde{\cal V}\right)
{\frac {B}{a_1t^2}}.\label{dB}
\end{equation}
Combining this with (\ref{dt2}), we obtain the solution:
\begin{equation}
B = {\frac {-t^2e^\rho}{\rho}}\,B_0,\label{B}
\end{equation}
with an arbitrary $B_0 =B_0(\zeta)$. 

Thus, we have completely determined the coframe (\ref{theta}) in terms of the
torsion 1-form (\ref{TdR}), the dual torsion (\ref{hT}) and the quadratic
torsion invariant (\ref{t2}). The spacetime interval (\ref{interval}) is then 
recovered straightforwardly as
\begin{equation}
ds^2 = -\,{\frac {(-t^2)e^{2\rho}}{\rho^2}}\,B_0^2d\zeta^2 
+ {\frac {a^2}{a_1^2}}{\frac {d\rho^2}{(-t^2)}}.\label{ds2}
\end{equation}
Without loss of generality, we can evidently put $B_0=1$ by a 
redefinition of the local coordinate $\zeta$.

In order to reveal the meaning of the integration constant $c_0$ in (\ref{t2}),
we notice that $\xi = \partial_\zeta$ is the Killing vector field for the 
line element (\ref{ds2}). As a result, we can verify that the energy 1-form 
\begin{equation}
\varepsilon = \xi^\alpha\,\Sigma_\alpha 
\end{equation}
is strongly conserved, $d\varepsilon = 0$. The first MAG 
field equation (\ref{first}) yields
\begin{equation}
\varepsilon = {\frac {a_1e^\rho}{2\rho}}\left[d(t^2) + (1 - 1/\rho)\,t^2d\rho
+ {\frac {a^2}{a_1^2}}(2\lambda - bR^2/2)\,d\rho\right]. 
\end{equation}
Using this, we can verify that $\varepsilon = dM$ with
\begin{equation}
M = {\frac {a_1t^2e^\rho}{2\rho}} + {\frac {a^2e^\rho}{4a_1b}}\Bigg\{
{\frac {[4b\lambda - (1 - 4a_2)2]}{a}}\,e^{-\rho}\,{\rm Ei}(\rho) 
 -\,a\rho + a - 2(1 - 4a_2)\Bigg\}.\label{M}
\end{equation}
In vacuum, $\Sigma_\alpha = 0$ and hence $\varepsilon = dM =0$. Thus, $M = M_0$
is constant for the vacuum solution obtained. Substituting (\ref{t2}) into
(\ref{M}), we find explicitly
\begin{equation}
c_0 = -\,{\frac {4a_1b}{a^2}}\,M_0.
\end{equation}
By construction, $M_0$ represents the total mass of the configuration, cf.
the discussion in \cite{col}.

\subsection{Torsion-degenerate spacetime}

Above, we have described a non-degenerate metric-affine spacetime, the
geometric properties of which are totally determined by the 2-dimensional
torsion. For completeness, we also need to analyse the case of degenerate
torsion with $t^2 =0$ everywhere. It is straightforward to see that the
first and the second MAG field equations (\ref{first}), (\ref{second}) yield
\begin{equation}
T^\alpha = 0,\qquad Q_{\alpha\beta} = 0,\qquad R = 2\sqrt{\frac \lambda b}.
\end{equation}
The resulting manifold is thus isometric to the 2-dimensional de Sitter space 
with vanishing torsion and nonmetricity. This result is analogous to the 
degenerate solutions of the Poincar\'e gauge model \cite{kat,sol,col,rev1}.

\section{Discussion and conclusion}\label{disc}

In this paper we have studied the quadratic MAG model (\ref{V}) in  
2-dimensional spacetime. We demonstrated that such a theory represents a new 
example of a completely integrable model of 2-dimensional gravity. This 
result was obtained by means of a direct extension of the general framework
which was developed earlier for the case of the Poincar\'e gauge models. Not
surprisingly, the form of the general vacuum solution resembles the solution 
of the quadratic Poincar\'e model. However, the old results are not recovered 
in the formal limit of the vanishing nonmetricity coupling constant $a_2
\rightarrow 0$. In particular, let us recall that it is possible to interpret
the solutions of the Poincar\'e model as 2-dimensional black holes. In our
current approach, the analysis of the possible black hole structure is related 
to the study of zeros of the metric coefficient $g_{\zeta\zeta} = t^2e^{2
\rho}/\rho^2$ for the general solution (\ref{t2}). For certain values of the 
coupling constants $(a_1, a_2, b, \lambda)$, the resulting geometry may indeed
display the black hole features similar to the black holes discovered in the 
quadratic Poincar\'e model. However, in general, the new solutions obtained 
are no black holes.

As compared to the Poincar\'e gauge case, qualitatively, the curvature and 
the torsion remain the basic elements of the theory. The scalar curvature
acts as one of the local coordinates of the spacetime manifold, whereas the
torsion defines a special coframe and thus provides the tool for constructing
the spacetime interval. In addition, one of our primary goals was to study
the specific role and place of the nonmetricity. Quite interestingly, it turns
out that the non-Riemannian sector of the MAG model is represented by the 
triplet structure $(T, Q, \Lambda)$ of the torsion and nonmetricity 1-forms. 
Such a triplet ansatz plays an important role in the 4-dimensional case 
\cite{eff}. The presence of the nonmetricity strongly modifies the vacuum
solution. In particular, it introduces more singularities into the Riemannian
geometry as compared to the effect of the torsion in the Poincar\'e model. The 
Riemannian curvature of the metric (\ref{ds2}) can be easily computed from the
corresponding Christoffel symbols and it reads explicitly
\begin{equation}
\widetilde{R} = R - {\frac {2a_1^2t^2}{a^2\rho^2}} - {\frac 2{ab}}
\,[4b\lambda - (bR)^2]\,(1 - 1/\rho).
\end{equation}

One may notice that the Lagrangian (\ref{V}) does not describe the most 
general MAG model in two dimensions. Indeed, since the nonmetricity has three 
irreducible parts, we can extend (\ref{V}) by including two more quadratic
$(Q\cdot Q)$ terms, and furthermore to add a nonmetricity times torsion
term of the form $(Q\cdot T)$. Such an extension, however, does not change 
our main result: The dynamics of the MAG fields remains qualitatively the 
same as in the minimal model (\ref{V}), and the generalized model is also
completely integrable. However, the inclusion of the nontrivial matter 
sources (scalar or spinor) as well as further extension of the model (\ref{V})
by adding the quadratic terms of $Z_{\alpha\beta}$ destroys the 
integrability, in general. 

The last but not least remark is about the origin of integrability property 
of the MAG model. There exists a systematic way of embedding the 2-dimensional 
(dilaton and Poincar\'e gauge) gravity into the class of so-called 
Poisson-Sigma models \cite{str,rev2}. The corresponding theoretical machinery 
provides an effective tool for demonstrating their complete integrability.
A preliminary analysis shows that a similar interpretation of the new MAG 
model as a Poisson-Sigma model seems also to be possible. 

\medskip
{\bf Acknowledgments}. The author is grateful to the Organizers of the 
workshop ``Gravity in two dimensions" for the invitation to the Erwin
Schr\"odinger Institute (Vienna) and for support. The discussions with the 
participants of the workshop are appreciated, with special thanks to Friedrich
Hehl for the helpful comments. This research was supported by the Deutsche 
Forschungsgemeinschaft (Bonn) with the grant HE~528/20-1.

\end{document}